\documentclass[amstex,thmsa,11pt,sw20bams]{article}%
\usepackage{amsmath}
\usepackage{amsfonts}
\usepackage[utf8]{inputenc}
\usepackage{graphicx}
\usepackage[francais]{babel}
\usepackage{amssymb}%
\providecommand{\U}[1]{\protect\rule{.1in}{.1in}}
\ifx\pdfoutput\relax\let\pdfoutput=\undefined\fi
\newcount\msipdfoutput
\ifx\pdfoutput\undefined\else
\ifcase\pdfoutput\else
\ifx\paperwidth\undefined\else
\ifdim\paperheight=0pt\relax\else\pdfpageheight\paperheight\fi
\ifdim\paperwidth=0pt\relax\else\pdfpagewidth\paperwidth\fi
\fi\fi\fi
\begin{document}

\author{Choulakian V., Universit\'{e} de Moncton, Canada
\and De Tibeiro J., Universit\'{e} de Moncton, Canada
\and Sarnacchiaro P., The University of Naples Federico II, Italy}
\title{On the choice of weights in aggregate compositional data analysis}
\date{January 2023}
\maketitle

\begin{abstract}
In this paper, we distinguish between two kinds of compositional data sets:
elementary and aggregate. This fact will help us to decide the choice of the
weights to use in log interaction analysis of aggregate compositional vectors.
We show that in the aggregate case, the underlying given data form a paired
data sets composed of responses and qualitative covariates; this fact helps us
to propose two approaches for analysis-visualization of data named log
interaction of aggregates and aggregate of log interactions. Furthermore, we
also show the first-order approximation of log interaction of a cell for
different choices of the row and column weights.

Key words: elementary and aggregate compositional vectors; log interaction of
aggregates; aggregate of log interactions; taxicab singular value
decomposition; row and column weights.

AMS 2010 subject classifications: 62H25, 62H30

\end{abstract}

\section{\textbf{Introduction}}

We start by comparing the structure of two compositional data sets, accessible
on the internet, which have been analyzed-visualized by centered clr
transformation, that we name log interaction approach.

First, the archeological CUPS compositional data set $\mathbf{X}=(x_{ij}),$ of
size $47\times11$, for $i=1,...,I$ rows and $j=1,...,J$ columns, found in
Greenacre and Lewi (2009); where each of the 47 cups is described by the
percentages by weight of 11 chemical elements. This data set has been also
considered by Allard et al. (2020).

Second, the FOOD compositional data set $(t_{kj}),$ of size $25\times9$, for
$k=1,...,K$ rows and $j=1,...,J$ columns, found in Pawlowsky-Glahn and Egozcue
(2011). These data are percentages of consumption of 9 different kinds of food
in 25 countries in Europe in the early eighties. This data set has been also
considered by Choulakian et al. (2023).

The CUPS and FOOD data sets are compositional in nature, but of different
structure: this is the main topic of this paper that we want to highlight and
discuss. Additionally, our discussion will lead further on to the choice of
the weights used in an analysis.

\subsection{Remarks}

a) Note that there are three different indices $(i,j,k)$ in $\mathbf{X=}%
(x_{ij})$ or $(t_{kj})$: the index $j$ being common to both.

b) We name the structure of CUPS data \textit{elementary}; while the structure
of FOOD data \textit{aggregate}, because it is obtained from two paired data
sets $\mathbf{X}=(x_{ij})$ for $i=1,...,I$ and $j=1,...,J$; and $\mathbf{Z}%
=(z_{ik})$ for $i=1,...,I$ and $k=1,...,K$, where $z_{ik}=0\ $or $1$:
$z_{ik}=1$ if observation $i$ belongs to the country $k$ and $z_{ik}=0$ if
observation $i$ does not belong to the country $k$. In FOOD data
$\mathbf{T}_{w_{i}^{I}}=\mathbf{Z}^{^{\prime}}\mathbf{D}_{I}\mathbf{X=}%
(t_{kj})$, where $\mathbf{D}_{I}=Diag(w_{i}^{I})$ is a diagonal metric matrix
with $w_{i}^{I}>0$ and $I$ is unknown. Often, the columns of \textbf{X}
represent the response variables, while the columns of \textbf{Z} represent
the categories of one or more qualitative covariates: in the case of FOOD
data, the columns of \textbf{Z} represent the categories of the qualitative
covariate country. The distinction between the two types, elementary and
aggregate of a compositional data set is implicit in the discussion of the
FOOD data by Pawlowsky-Glahn and Egozcue (2011, section 2).

c) Similar to \textbf{D}$_{I}=Diag(w_{i}^{I}),$ we define two other metric
matrices: \textbf{D}$_{J}=Diag(w_{j}^{J})$ is a diagonal metric matrix with
$w_{j}^{J}>0$ for the rows of \textbf{X, }and \textbf{D}$_{K}=Diag(w_{k}^{K})$
is a diagonal metric matrix with $w_{k}^{K}>0$ for the rows of \textbf{Z.}

\subsection{Aims and organization}

The aim of this paper is the study of a compositional data set composed of the
paired sets \textbf{X} and \textbf{Z} in two different approaches: one of them
being the aggregate matrix $\mathbf{T}_{w_{i}^{I}}=\mathbf{Z}^{^{\prime}%
}\mathbf{D}_{I}\mathbf{X};$ the other one being new to our knowledge, named
aggregate of log interaction approach. That is, we present two approaches that
incorporate the 0/1 covariates \textbf{Z} into \textbf{X}. The recent paper by
D'Ambra et al. (2020) discuss the analysis of paired sets \textbf{X} and
\textbf{Z }by correspondence analysis related method.

This paper is organized as follows: Section 2 presents the paired data sets
\textbf{X} and \textbf{Z }that we will use in this paper; section 3 reviews
the common well known log interaction analysis of \textbf{T}; section 4
presents the new approach named aggregate of log interactions; section 5
discusses the choice of weights; section 6 presents the analysis of the data
set by three distinct methods; we conclude in section 7.

\section{Household expenditure paired data sets}

The paired data sets are \textbf{X} of size $166\times9$ and \textbf{Z} \ of
size $166\times11$ and are available from the second author.

$I=166$ is the sample size of families observed in Napoli Italy. $J=9$
represents 9 essential household expenditure items: \textit{grocery shopping,
clothing, house, fuel, health, holydays, car maintenance, insurance, household
electrical appliances}. So $x_{ij}$ represents monthly amount (in euros) spent
by the head of the $i$th family on the $j$th household item. Note that
$x_{ij}>0$, so log $x_{ij}$ is finite.

Each head of the 166 families is characterized by $Q=4$ qualitative variables:
\textit{Sex (male (M), female (F)), Age (Less than 31 years = "Age1", between
31 and 45 years = "Age2", more than 45 years = "Age3"), Income (low = "Inc1",
average = "Inc2", high = "Inc3"), Education (low education = "Ed1", average
education = "Ed2", high education = "Ed3")}. So $z_{ik}$ represents the $k$th
caracteristic of the head of the $i$th family; for instance, $z_{i2}$ is the
indicator (0/1) dummy variable whether the head of the $i$th family is a
female, $z_{i6}$ has the value (0 or 1) meaning whether the head of the $i$th
family has a low income level (Inc1). The marginals of the rows of \textbf{Z},
$z_{i+}=\sum_{k=1}^{11}z_{ik}=Q=4$, because there are 4 qualitative
covariates. The marginals of the columns of \textbf{Z}, $z_{+k}=\sum
_{i=1}^{166}z_{ik}$ are:%
\begin{equation}
(117,49,57,41,68,48,54,64,45,52,69), \tag{1}%
\end{equation}
where we see $\sum_{k=1}^{2}z_{+k}=\sum_{k=3}^{5}z_{+k}=\sum_{k=6}^{8}%
z_{+k}=\sum_{k=9}^{11}z_{+k}=166;$ that is the 166 heads of the families are
partitioned in $Q=4$ different ways.

\section{Aggregate compositional data analysis}

Assume $\mathbf{D}_{I}=(1/I=1/166=z_{i+}/z_{++}=Q/(QI)$, where $z_{i+}%
=\sum_{k=1}^{K}z_{ik}=Q$ and $z_{++}=\sum_{i=1}^{I}\sum_{k=1}^{K}z_{ik}=QI.$
Then the aggregate compositional data is $\mathbf{T}=\mathbf{Z}^{^{\prime}%
}\mathbf{X/}I$ of size $K\times J=11\times9;$ and $It_{kj}=\sum_{i=1}%
^{I}z_{ik}x_{ij}$ represents the total amount spent on household item $j$ by
all heads of the families having the characteristic $k$. By (1), we see that
\textbf{T} is made up of $Q=4$ vertically juxtaposed aggregate comositional
data blocks: $\mathbf{T}_{sex},\mathbf{T}_{age},\mathbf{T}_{income}%
,\mathbf{T}_{education}.$

Assume $t_{kj}>0$ and define $G_{kj}=\log(p_{kj}^{KJ})$ where $p_{kj}%
^{KJ}=t_{kj}/t_{++}$ and $t_{++}=\sum_{k,j}t_{kj},$ then ($w_{k}^{K}$,
$w_{j}^{J})$ weighted log interaction between the $k$th caracteristic and the
$j$th item is,%

\begin{equation}
\lambda_{kj}^{KJ}=G_{kj}-G_{k+}-G_{+j}+G_{++}, \tag{2}%
\end{equation}
where%

\[
G_{k+}=\sum_{j=1}^{J}G_{kj}w_{j}^{J},
\]
\[
G_{+j}=\sum_{k=1}^{K}G_{kj}w_{k}^{K},
\]
\[
G_{++}=\sum_{j=1}^{J}\sum_{k=1}^{K}G_{kj}w_{j}^{J}w_{k}^{K};
\]
$w_{k}^{K}>0$ and $w_{j}^{J}>0$ are a priori fixed or data dependent
probability weights, satisfying $\sum_{j=1}^{J}w_{j}^{J}=\sum_{k=1}^{K}%
w_{k}^{K}\ =1$.

We note that $w_{j}^{J}w_{k}^{K}\lambda_{kj}^{KJ}$ for $k=1,...,K$ and
$j=1,...,J$ are row and column centered (double centered); that is%

\begin{align}
0  &  =w_{k}^{K}\sum_{j=1}^{J}w_{j}^{J}\lambda_{kj}^{KJ}\text{ \ for
\ \ }k=1,...,K\nonumber\\
&  =w_{j}^{J}\sum_{k=1}^{K}w_{k}^{K}\lambda_{kj}^{KJ}\text{\ \ \ for
\ }j=1,...,J. \tag{3}%
\end{align}

Here we cite two important results concerning (2); see (Choulakian 2022, or
Choulakian et al. 2023). First, the log interaction terms in (2) are scale
invariant for fixed a priori weights ($w_{k}^{K},w_{j}^{J})$. Note that in
(2), $p_{kj}^{KJ}$ depends on $t_{kj},$ $p_{kj}^{KJ}=t_{kj}/\sum_{k,j}%
t_{kj}=t_{kj}/t_{++},$ where $t_{++}=\frac{Qx_{++}}{I}$. To emphasize this
dependence, we express the interaction index (2) by $\lambda_{kj}^{KJ}%
(t_{kj})=\lambda(p_{kj}^{KJ},w_{k}^{K},w_{j}^{J})$. Following Yule (1912) and
Goodman (1996), we state the following\bigskip

\textbf{Definition 1}: An interaction index $\lambda_{ij}(n_{ij})$ is scale
invariant if $\lambda_{ij}(n_{ij})=\lambda_{ij}(a_{i}n_{ij}b_{j})$ for scales
$a_{i}>0$ and $b_{j}>0$.\bigskip

\textbf{Lemma 1}: a) The interaction index $\lambda_{kj}^{KJ}$ in (2) is scale
invariant.\bigskip

b) Assuming $t_{kj}>0,$ we have to a first-order approximation,
\[
\lambda_{kj}^{KJ}\approx\widetilde{\lambda}_{kj}^{KJ}=(\frac{p_{kj}^{KJ}%
}{w_{k}^{K}w_{j}^{J}}-\frac{p_{k+}^{KJ}}{w_{k}^{K}}-\frac{p_{+j}^{KJ}}%
{w_{j}^{J}}+1).
\]
\textbf{Corollary 1}:

a) $w_{k}^{K}w_{j}^{J}\widetilde{\lambda}_{kj}^{KJ}$ for $k=1,...,K$ and
$j=1,...,J$ are row and column centered; that is, they satisfy the two
equations in (3), where $\lambda_{kj}^{KJ}$ is replaced by $\widetilde{\lambda
}_{kj}^{KJ}.$

b) If $w_{k}^{K}=p_{k+}^{KJ}$ and $w_{j}^{J}=p_{+j}^{KJ},$ then $\lambda
_{kj}^{KJ}\approx\widetilde{\lambda}_{kj}^{KJ}=\frac{p_{kj}^{KJ}}{p_{k+}%
^{KJ}p_{+j}^{KJ}}-1,$ which is the correspondence analysis (CA)$\ $interaction
of the $k$th row and the $j$th column of \textbf{T}. That is, correspondence
analysis (CA) of \textbf{T} is a first-order approximation of the log
interaction analysis of \textbf{T}, a well-known result; for a review see
Choulakian et al. (2023).

\section{Aggregate of elementary log interactions analysis}

Consider the elementary compositional data set \textbf{X}. Assume $x_{ij}>0$
and define $G_{ij}=\log(p_{ij}^{IJ}),$ where $p_{ij}^{IJ}=x_{ij}/x_{++}$ and
$x_{++}=\sum_{i,j}x_{ij}.$ Then $(w_{i}^{I}$, $w_{j}^{J})$ weighted log
interaction between the $i$th individual (sometimes named sample) and the
$j$th column of \textbf{X} is,%

\begin{equation}
\lambda_{ij}^{IJ}=G_{ij}-G_{i+}-G_{+j}+G_{++}, \tag{4}%
\end{equation}
where

$G_{i+}=\sum_{j=1}^{J}G_{ij}w_{j}^{J},$ $G_{+j}=\sum_{i=1}^{I}G_{ij}w_{i}^{I}$
and $G_{++}=\sum_{j=1}^{J}\sum_{i=1}^{I}G_{ij}w_{j}^{J}w_{i}^{I}$; $w_{i}%
^{I}>0$ and $w_{j}^{J}>0,$ satisfying $\sum_{j=1}^{J}w_{j}^{J}=\sum_{i=1}%
^{I}w_{i}^{I}\ =1,$ are a priori fixed or data dependent probability weights.

We designate by $\Lambda^{IJ}=(\lambda_{ij}^{IJ})$ of size $I\times J$ the
matrix of elementary log interactions, and we consider the aggregate of the
elementary log interactions to be $\mathbf{A}_{w_{i}^{I}}=(\alpha
_{kj})=\mathbf{Z}^{^{\prime}}\mathbf{D}_{I}\Lambda^{IJ}.$

Using the following identities $x_{+j}=It_{+j}/Q,$ $x_{++}=It_{++}/Q$ and
$t_{kj}=\sum_{i=1}^{I}z_{ik}x_{ij}/I$, where $w_{i}^{I}=1/I$, and by Lemma 1b
applied to $\mathbf{A}_{w_{i}^{I}},$ we have the first-order approximation
$\widetilde{\alpha}_{kj},$
\begin{align}
\alpha_{kj}  &  =\sum_{i=1}^{I}z_{ik}w_{i}^{I}\lambda_{ij}^{IJ}\nonumber\\
&  \approx\sum_{i=1}^{I}z_{ik}\frac{1}{I}\widetilde{\lambda}_{ij}%
^{IJ}=\widetilde{\alpha}_{kj}\nonumber\\
&  =\sum_{i=1}^{I}z_{ik}\frac{1}{I}(\frac{p_{ij}^{IJ}}{\frac{1}{I}w_{j}^{J}%
}-\frac{p_{i+}^{IJ}}{\frac{1}{I}}-\frac{p_{+j}^{IJ}}{w_{j}^{J}}+1)\nonumber\\
&  =\frac{1}{x_{++}}(\sum_{i=1}^{I}z_{ik}\frac{x_{ij}}{w_{j}^{J}}-\sum
_{i=1}^{I}z_{ik}x_{i+}-\frac{x_{+j}}{w_{j}^{J}}\sum_{i=1}^{I}z_{ik}\frac{1}%
{I})+\sum_{i=1}^{I}z_{ik}\frac{1}{I}\nonumber\\
&  =\frac{Q}{It_{++}}(\frac{I}{w_{j}^{J}}t_{kj}-It_{k+}-\frac{It_{+j}}%
{Qw_{j}^{J}}z_{+k}\frac{1}{I})+z_{+k}\frac{1}{I}\nonumber\\
&  =Q(\frac{p_{kj}^{KJ}}{w_{j}^{J}}-p_{k+}^{KJ})-(\frac{p_{+j}^{KJ}}{w_{j}%
^{J}}-1)\frac{z_{+k}}{I}. \tag{5}%
\end{align}
\textbf{Corollary 2}:

a) $(\frac{1}{I}w_{j}^{J}\lambda_{ij}^{IJ})$ and $(\frac{1}{I}w_{j}%
^{J}\widetilde{\lambda}_{ij}^{IJ})$ for $i=1,...,I$ and $j=1,...,J$ are row
and column centered.

b) $(\frac{1}{K}w_{j}^{J}\alpha_{kj})$ and $(\frac{1}{K}w_{j}^{J}%
\widetilde{\alpha}_{kj})$ for $k=1,...,K$ and $j=1,...,J$ are row and column centered.

c) If $w_{j}^{J}=\sum_{i=1}^{I}x_{ij}/x_{++}=p_{+j}^{IJ}=p_{+j}^{KJ},$ then by
(5) we have%
\[
\alpha_{kj}\approx Q(\frac{p_{kj}^{KJ}}{p_{+j}^{KJ}}-p_{k+}^{KJ});
\]
which is $Q$ times the centering used in nonsymmetric correspondence analysis
(NSCA) of \textbf{T,} with weights $w_{k}^{K}=1/K$ and $w_{j}^{J}=p_{+j}%
^{IJ}=p_{+j}^{KJ};$see Lauro and D$^{\prime}$ambra (1984).\textbf{ }That is,
NSCA of \textbf{T} is a first-order approximation of the $Q\mathbf{A}%
_{w_{i}^{I}}$ with $w_{i}^{I}=1/I$.

\section{The choice of weights}

Benz\'{e}cri (1973, Vol.1, p. 31-32) emphasized the importance of the row and
column weights or metrics in multidimensional data analysis. Choulakian et al.
(2023) present a review of literature concerning the choice of weights in log
interaction analysis. We have three cases:

a) Consider the elementary compositional data set $\mathbf{X}=(x_{ij})$.
Aitchison (1983,1986,1994,1997) used the uniform weights $w_{i}^{I}=1/I$,
$w_{j}^{J}=1/J$; and this has been followed up. Its first-order approximation
is, by Lemma 1b,%
\[
\lambda_{ij}^{IJ}\approx\widetilde{\lambda}_{ij}^{IJ}=(IJp_{ij}^{IJ}%
-Ip_{i+}^{IJ}-Jp_{+j}^{IJ}+1).
\]

b) Consider the aggregate compositional data set $\mathbf{T}=(t_{kj})$. We
suggest the use of the following weights $w_{k}^{K}=\sum_{i=1}^{I}z_{ik}%
/\sum_{i=1}^{I}\sum_{k=1}^{K}z_{ik}$ $=z_{+k}/z_{++}$ \ and $w_{j}^{J}=1/J,$
essentially for the following two reasons. First, by looking at (1), we see
that the sample size of each category $k$ is different, and this should be
taken into account. For example, the number of males observed is $117,$ while
the number of females observed is $49$, as displayed in (1). Second, by Lemma
1a, the aggregate vector of the $k$th category is equivalent to its average.
By Lemma 1b, the first-order approximation of $\lambda_{kj}^{KJ}$ is,%
\[
\lambda_{kj}^{KJ}\approx\widetilde{\lambda}_{kj}^{KJ}=(\frac{Jp_{kj}^{KJ}%
}{z_{+k}/z_{++}}-\frac{p_{k+}^{KJ}}{z_{+k}/z_{++}}-Jp_{+j}^{KJ}+1).
\]

c) Consider the aggregate of the elementary log interactions $\mathbf{A}%
=(a_{kj})=\mathbf{Z}^{^{\prime}}\Lambda^{IJ}/I;$ note that here $\mathbf{D}%
_{I}=(1/I=1/166=z_{i+}/z_{++}=Q/(QI)$. We suggest the use of the following
weights $w_{k}^{K}=1/K$, $w_{j}^{J}=1/J;$ because the effect of the sample
size of category $k$ has been taken into account by the summation of the log
interactions in the category $k$. The first-order approximation of $a_{kj}$
is, by (5),%
\begin{equation}
a_{kj}\approx\widetilde{a}_{kj}=Q(Jp_{kj}^{KJ}-p_{k+}^{KJ})-(Jp_{+j}%
^{KJ}-1)\frac{z_{+k}}{I}. \tag{6}%
\end{equation}

\section{Data analysis}

For data analysis purposes, we shall apply taxicab singular value
decomposition (TSVD) to a matrix, see Choulakian (2006, 2016), and its related
statistics as described in the appendixes 1, 2 and 3.

\begin{figure}[h]
\label{fig:MagEngT}
\centering 
\includegraphics[scale=0.55]{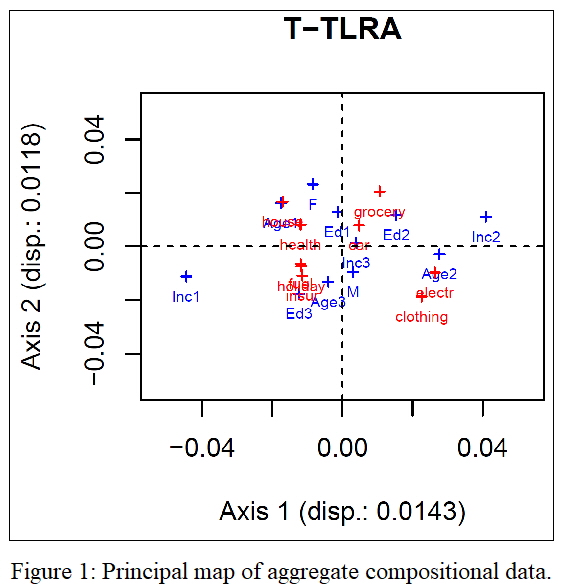}  
\end{figure}

Figures 1 and 2 represent the taxicab principal maps obtained from TSVD of
\textbf{T} and \textbf{A}: Which one is \textquotedblright
preferable\textquotedblright? Figure 2 for the following reasons: First, the
QSR index shows that the decomposition of the \textbf{A} matrix is preferable
to the decomposition of the \textbf{T} matrix: For the \textbf{A} matrix,
\[
QSR_{1}+QSR_{2}=54.8+72.5=127.3,
\]
which is significantly greater than the corresponding values of the \textbf{T}
matrix
\[
QSR_{1}+QSR_{2}=55.2+59.9=115.1.
\]
Second, the first two taxicab principal dispersion measures (0.633 and 0.619)
of \textbf{A} are clearly separated from the 3rd and 4rth taxicab principal
dispersion measures (0.361 and 0.264). Third, the interpretation of the
principal map obtained from the \textbf{A} matrix is easier, because of the
uniform weightings of the K categories and the J items.

\begin{figure}[h]
\label{fig:MagEngT}
\centering 
\includegraphics[scale=0.55]{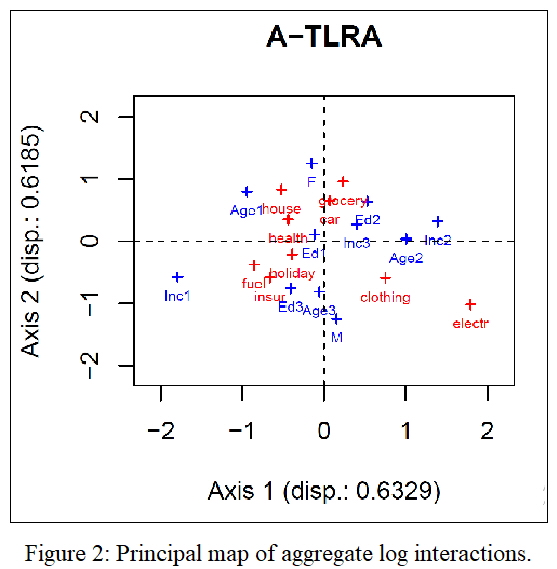}  
\end{figure}

In Figure 2, the first taxicab principal dimension opposes (\textit{Inc1} and
\textit{Age1}) associated with (\textit{fuel} and \textit{Insur}) to
(\textit{Inc2} and \textit{age2}) associated with (\textit{electr} and
\textit{clothing}). The second taxicab principal dimension opposes
(\textit{male}, \textit{Age3} and \textit{Ed3}) associated with
(\textit{electr, clothing, insurance}) to (\textit{female, Age1} and
\textit{Ed2}) associated with (\textit{grocery, house, car}).

Now we ask the question: How about the 1st-order approximation of \textbf{A}
as derived in equation (6)? Figure 3 shows its taxicab principal map, where we
see clearly a reversal of the first two taxicab principal axes of the
\textbf{A} approach. The map is clearer: In the first quadrant are found
(I\textit{nc1,Age1} and \textit{Ed1, female}) associated with
(\textit{housing} and \textit{grocery}); in the second quadrant are found
(\textit{Inc3,Age3} and \textit{Ed3}) associated with (\textit{insurance,
holiday, car, healthcare}); in the third quadrant are found (\textit{male})
associated with (\textit{electrical appliances}); in the fourth quadrant are
found (\textit{Inc2,Age2} and \textit{Ed2}) associated with (\textit{clothing}%
). Furthermore, Figure 3 seems preferable to Figures 1 and 2, based on the the
comparison of the $QSR_{1}$ and $QSR_{2}$ values, where
\[
QSR_{1}+QSR_{2}=73.9+59.4=133.3,
\]
which is greater than the corresponding values of \textbf{T} ($115.1)$ and
\textbf{A} ($127.3)$. Now we interpret the QSR values of the quadrants for the
first and second principal axes. Any quadrant QSR index can be interpreted as
an association index given that $-1\leq QSR\leq1$ or $-100\%\leq
QSR\leq100\%,$ see Choulakian (2021) and Choulakian and Abou-Samra (2020).

The first principal row and column axes ($\mathbf{u}_{1},\mathbf{v}_{1})$
divide the interaction matrix (6) into four principal quadrants where two
quadrants $(V_{+}U_{+},V_{-}U_{-})$ are positively associated and two
quadrants are negatively associated $(V_{-}U_{+},V_{+}U_{-}).$ We see that
$QSR_{1}(V_{-}U_{-})=88.2\%$ is quite high; which means that the variables
\[
V_{-}=\{Inc3,\ Age3,\ Ed3,male\}
\]
are quite highly positively associated with%
\[
U_{-}=\{insurance,electrical\ appliances,\ healthcare,holiday\}.
\]

Similarly, we see that $QSR_{1}(V_{-}U_{+})=-87.1\%$ is quite high in absolute
value; which means that the variables
\[
V_{-}=\{Inc3,\ Age3,\ Ed3,male\}
\]
are quite highly negatively associated with%
\[
U_{+}=\{grocery,housing,clothing\}.
\]

A similar interpretation can be provided on the intensity of the four
principal quadrants designed by the second principal axis. We see that
$QSR_{2}(V_{-}U_{-})=76.8\%$ is high; which means that the variables
\[
V_{-}=\{Inc2,\ Age2,\ Ed2,male\}
\]
are highly positively associated with%
\[
U_{-}=\{clothing,electrical\ appliances\}.
\]

Similarly, we see that $QSR_{2}(V_{-}U_{+})=-71.3\%$ is high in absolute
value; which means that the variables
\[
V_{-}=\{Inc2,\ Age2,\ Ed2,male\}
\]
are highly negatively associated with%
\[
U_{+}=\{housing,holidays,fuel\}.
\]

\begin{tabular}
[c]{lllll}%
\multicolumn{5}{l}{\textbf{Table 1: QSR (\%) for the first 4 dimensions.}%
}\\\hline\hline
&  & T-TLRA &  & \\
Axis $\alpha$ & $QSR_{\alpha}(V_{+}U_{+},V_{-}U_{-})$ & $QSR_{\alpha}%
(V_{-}U_{+},V_{+}U_{-})$ & $QSR_{\alpha}$ & $\delta_{\alpha}$\\\hline
1 & (61.6,\textbf{\ }42.6) & (-61.2,\ -60.9\textbf{)} & \textbf{55.2} &
\textbf{0.0143}\\
2 & (68.7,\ 52.4\textbf{)} & (-67.5\textbf{,\ }-54.4) & \textbf{59.9} &
\textbf{0.0118}\\
3 & (70.4,\ 63.3) & (-71.9, -56.6) & 65.0 & 0.0106\\
4 & (87.7, 74.9) & (-100, -66.2) & 80.2 & 0.0092\\\hline\hline
&  & A-TLRA &  & \\
Axis $\alpha$ & $QSR_{\alpha}(V_{+}U_{+},V_{-}U_{-})$ & $QSR_{\alpha}%
(V_{-}U_{+},V_{+}U_{-})$ & $QSR_{\alpha}$ & $\delta_{\alpha}$\\\hline
1 & (54.5,\ 48.3) & (-47.7,\ -77.0) & \textbf{54.8} & \textbf{0.633}\\
2 & (70.6,\ 78.5) & (-92.5,\ -57.3) & \textbf{72.5} & \textbf{0.619}\\
3 & (73.0,\ 60.4) & (-56.2, -65.4) & 63.2 & 0.361\\
4 & (89.1, 42.6) & (-71.2, -69.1) & 63.2 & 0.264\\\hline\hline
&  & A-1st order TSVD &  & \\
$\alpha$ & $QSR_{\alpha}(V_{+}U_{+},V_{-}U_{-})$ & $QSR_{\alpha}(V_{-}%
U_{+},V_{+}U_{-})$ & $QSR_{\alpha}$ & $\delta_{\alpha}$\\\hline
1 & (67.3\textbf{,\ 88.2}) & (\textbf{-87.1},\ -60.9\textbf{)} & \textbf{73.9}
& \textbf{0.939}\\
2 & (47.0,\ \textbf{76.8)} & (\textbf{-71.3,\ -}52.5\textbf{)} & \textbf{59.4}
& \textbf{0.448}\\
3 & (57.5,\ 82.0) & (-72.0, -66.7) & 68.4 & 0.391\\
4 & (79.5, 88.3) & (-58.3, -94.3) & 77.4 & 0.297\\\hline
\end{tabular}

\begin{figure}[h]
\label{fig:MagEngT}
\centering 
\includegraphics[scale=0.55]{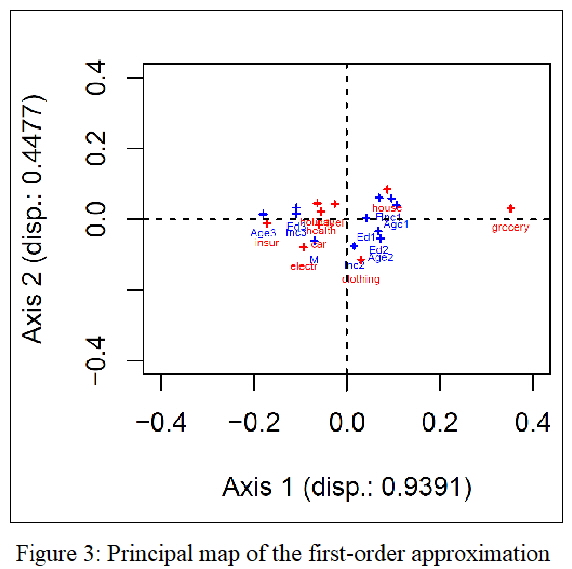}  
\end{figure}

\section{Conclusion}

We conclude by summarizing the topics discussed in this paper.

First, we emphasize the fact that the inclusion of the row and column weights
in logarithmic approach of analyzing-visualizing a compositional data set
renders it quite versatile.

Second, we distinguished between elementary (\textbf{X}) and aggregate
compositional data sets (\textbf{T); T }is built from the paired data sets
(\textbf{X}, \textbf{Z}), where \textbf{X} is the response data set and
\textbf{Z} the indicator matrix with elements (0,1) of \textbf{Q} qualitative covariates.

Third, we proposed to analyze-visualize the paired data sets (\textbf{X},
\textbf{Z}) by two approaches named, aggregate compositional data (\textbf{T})
and aggregate log interactions (\textbf{A}).

Fourth, we presented the first-order approximation to the log interaction of a
cell under different weighting schemes; and we think these can be used in case
the original data contain zero cells. This needs further study.\bigskip
\begin{verbatim}
Acknowledgements: Choulakian's research has been supported by NSERC of Canada.
\end{verbatim}

\bigskip
\begin{verbatim}
Appendix 1: An overview of taxicab singular value decomposition
\end{verbatim}

Consider a matrix $\mathbf{X}$\ of size $I\times J$ and $rank(\mathbf{X}%
)=k$\textbf{.} Taxicab singular value decomposition (TSVD) of \textbf{X} is a
decomposition similar to SVD of \textbf{X}, see Choulakian (2006, 2016).

For a vector $\mathbf{u}=\mathbf{(}u_{i}),$ its taxicab or L$_{1}$ norm is
$\left\vert \left\vert \mathbf{u}\right\vert \right\vert _{1}=\sum_{i}%
|u_{i}|,$ the Euclidean or L$_{2}$ norm is $\left\vert \left\vert
\mathbf{u}\right\vert \right\vert _{2}=(\sum_{i}|u_{i}|^{2})^{1/2}$ and the
L$_{\infty}$ norm is $\left\vert \left\vert \mathbf{u}\right\vert \right\vert
_{\infty}=\max_{i}\ |u_{i}|$.

In TSVD the calculation of the dispersion measures $(\delta_{\alpha})$,
principal axes ($\mathbf{u}_{\alpha},\mathbf{v}_{\alpha})$ and principal
scores $(\mathbf{a}_{\alpha},\mathbf{b}_{\alpha})$ for $\alpha=1,...,k$ is
done in a stepwise manner. We put $\mathbf{X}_{1}=\mathbf{X}=(x_{ij})$ and
$\mathbf{X_{\alpha}}$ be the residual matrix at the $\alpha$-th iteration for
$\alpha=1,...,k$.

The variational definitions of the TSVD at the $\alpha$-th iteration are%

\begin{align}
\delta_{\alpha}  &  =\max_{\mathbf{u\in%
\mathbb{R}
}^{J}}\frac{\left\vert \left\vert \mathbf{X_{\alpha}u}\right\vert \right\vert
_{1}}{\left\vert \left\vert \mathbf{u}\right\vert \right\vert _{\infty}}%
=\max_{\mathbf{v\in%
\mathbb{R}
}^{I}}\ \frac{\left\vert \left\vert \mathbf{X_{\alpha}^{\prime}v}\right\vert
\right\vert _{1}}{\left\vert \left\vert \mathbf{v}\right\vert \right\vert
_{\infty}}=\max_{\mathbf{u\in%
\mathbb{R}
}^{J},\mathbf{v\in%
\mathbb{R}
}^{I}}\frac{\mathbf{v}^{\prime}\mathbf{X_{\alpha}u}}{\left\vert \left\vert
\mathbf{u}\right\vert \right\vert _{\infty}\left\vert \left\vert
\mathbf{v}\right\vert \right\vert _{\infty}},\tag{A1}\\
&  =\max\ ||\mathbf{X_{\alpha}u||}_{1}\ \ \text{subject to }\mathbf{u}%
\in\left\{  -1,+1\right\}  ^{J},\nonumber\\
&  =\max\ ||\mathbf{X_{\alpha}^{\prime}v||}_{1}\ \ \text{subject to
}\mathbf{v}\in\left\{  -1,+1\right\}  ^{I},\nonumber\\
&  =\max\mathbf{v}^{\prime}\mathbf{X_{\alpha}u}\text{ \ subject to
\ }\mathbf{u}\in\left\{  -1,+1\right\}  ^{J},\mathbf{v}\in\left\{
-1,+1\right\}  ^{I}.\nonumber
\end{align}
The $\alpha$-th principal axes are%
\begin{equation}
\mathbf{u}_{\alpha}\ =\arg\max_{\mathbf{u}\in\left\{  -1,+1\right\}  ^{J}%
}\left\vert \left\vert \mathbf{X_{\alpha}u}\right\vert \right\vert _{1}\text{
\ \ and \ \ }\mathbf{v}_{\alpha}\ =\arg\max_{\mathbf{v}\in\left\{
-1,+1\right\}  ^{I}}\left\vert \left\vert \mathbf{X_{\alpha}v}\right\vert
\right\vert _{1}\text{,} \tag{A2}%
\end{equation}
and the $\alpha$-th principal projections of the rows and the columns are
\begin{equation}
\mathbf{a}_{\alpha}=\mathbf{X_{\alpha}u}_{\alpha}\text{ \ and \ }%
\mathbf{b}_{\alpha}=\mathbf{X_{\alpha}^{\prime}v}_{\alpha}. \tag{A3}%
\end{equation}
Furthermore, the following relations are also useful%
\begin{equation}
\mathbf{u}_{\alpha}=sign(\mathbf{b}_{\alpha})\text{ \ and \ }\mathbf{v}%
_{\alpha}=sign(\mathbf{a}_{\alpha}), \tag{A4}%
\end{equation}
where $sign(.)$ is the coordinatewise sign function, $sign(x)=1$ \ if \ $x>0,$
\ and \ $sign(x)=-1$ \ if \ $x\leq0.$

The $\alpha$-th taxicab dispersion measure $\delta_{\alpha}$ can be
represented in many different ways%
\begin{align}
\delta_{\alpha}\  &  =\left\vert \left\vert \mathbf{X_{\alpha}u}_{\alpha
}\right\vert \right\vert _{1}=\left\vert \left\vert \mathbf{a}_{\alpha
}\right\vert \right\vert _{1}=\mathbf{a}_{\alpha}^{\prime}\mathbf{v}_{\alpha
},\tag{A5}\\
&  =\left\vert \left\vert \mathbf{X_{\alpha}^{\prime}v}_{\alpha}\right\vert
\right\vert _{1}=\left\vert \left\vert \mathbf{b}_{\alpha}\right\vert
\right\vert _{1}=\mathbf{b}_{\alpha}^{\prime}\mathbf{u}_{\alpha}\nonumber\\
&  =\mathbf{v}_{\alpha}{}^{\prime}\mathbf{X_{\alpha}u}_{\alpha}%
=sign(\mathbf{a}_{\alpha})^{\prime}\mathbf{X_{\alpha}}sign(\mathbf{b}_{\alpha
}).\nonumber
\end{align}
The $(\alpha+1)$-th residual matrix is
\begin{equation}
\mathbf{X_{\alpha+1}}=\mathbf{X_{\alpha}-a}_{\alpha}\mathbf{b}_{\alpha
}^{\prime}/\delta_{\alpha}. \tag{A6}%
\end{equation}
An interpretation of the term $\mathbf{a}_{\alpha}\mathbf{b}_{\alpha}^{\prime
}/\delta_{\alpha}$ in (A6) is that, it represents the best rank-1
approximation of the residual matrix $\mathbf{X_{\alpha}}$, in the sense of
the taxicab matrix norm (A1).

Thus TSVD of $\mathbf{X}$ corresponds to the bilinear decomposition%

\begin{equation}
x_{ij}=\sum_{\alpha=1}^{k}a_{\alpha}(i)b_{\alpha}(j)/\delta_{\alpha}, \tag{A7}%
\end{equation}
a decomposition similar to SVD, but where the vectors $(\mathbf{a}_{\alpha
},\mathbf{b}_{\alpha})$ for $\alpha=1,...,k$ are conjugate, that is%
\begin{align}
\mathbf{a}_{\alpha}^{\prime}\mathbf{v}_{\beta}  &  =\mathbf{a}_{\alpha
}^{\prime}sign(\mathbf{a}_{\beta})\tag{A8}\\
&  =\mathbf{b}_{\alpha}^{\prime}\mathbf{u}_{\beta}=\mathbf{b}_{\alpha}%
^{\prime}sign(\mathbf{b}_{\beta})\nonumber\\
&  =0\text{ for }\alpha\geq\beta+1.\nonumber
\end{align}

\bigskip
\begin{verbatim}
Appendix 2: Taxicab matrix factorization with weights
\end{verbatim}

Let $\mathbf{Y}=(y_{ij}\mathbf{)}$ be a double-centered matrix with respect to
row and column weights ($m_{i}^{r},m_{j}^{c});$ \textbf{Y} can be any one of
the matrices \textbf{T, A}, discussed in the text. Then the taxicab
factorization of $\mathbf{Y}=(y_{ij}\mathbf{)}$ is done in three steps:

\textit{Step 1}: We double-center $\mathbf{Y}=(\tau_{ij}\mathbf{)}$%
\begin{equation}
X(i,j)=y_{ij}m_{i}^{r}m_{j}^{c} \tag{A9}%
\end{equation}

That is
\begin{align*}
0  &  =\sum_{i=1}^{I}X(i,j)\\
&  =\sum_{j=1}^{J}X(i,j).
\end{align*}
This double-centering step is necessary to have the important basic equations
(A14,A15) on which the QSR index is based.

\textit{Step 2}: Calculate TSVD of $\mathbf{X}=(y_{ij}m_{i}^{r}m_{j}%
^{c}\mathbf{)}$ as described in Appendix 1%
\begin{equation}
y_{ij}m_{i}^{r}m_{j}^{c}=\sum_{\alpha=1}^{k}a_{\alpha}(i)b_{\alpha}%
(j)/\delta_{\alpha}. \tag{A10}%
\end{equation}
We name $(a_{\alpha}(i),b_{\alpha}(j))$ taxicab contribution scores because
they satisfy following (A5)%
\begin{equation}
\delta_{\alpha}=\sum_{i=1}^{I}|a_{\alpha}(i)|=\sum_{j=1}^{J}|b_{\alpha
}(j)|\text{\ \ \ for }\alpha=1,...,k. \tag{A11}%
\end{equation}
Furthermore, they are centered following Step 1%

\begin{equation}
0=\sum_{i=1}^{I}a_{\alpha}(i)=\sum_{j=1}^{J}b_{\alpha}%
(j))\ \ \text{\ \ for\ \ \ \ \ }\alpha=1,...,k; \tag{A12}%
\end{equation}
And they are conjugate ( in TSVD conjugacy replaces orthogonality in SVD)%
\begin{equation}
0=\sum_{i=1}^{I}a_{\alpha}(i)\ sign(a_{\beta}(i))=\sum_{j=1}^{J}b_{\alpha
}(j)\ sign(b_{\beta}(j))\ \text{\ for\ \ \ \ }\alpha>\beta. \tag{A13}%
\end{equation}
Let $S=\left\{  i:a_{\alpha}(i)\geq0\right\}  $ and $T=\left\{  j:b_{\alpha
}(j)\geq0\right\}  $, so that at iteration $\alpha,$ $S\cup\overline{S}=I$ is
an optimal partition of $I$ and $T\cup\overline{T}=J$ is an optimal partition
of $J.$ Besides (A11), the taxicab dispersion $\delta_{\alpha}$ will
additionally satisfy the following useful equations:%
\begin{align}
\delta_{\alpha}/2  &  =\sum_{i\in S}a_{\alpha}(i)=-\sum_{i\in\overline{S}%
}a_{\alpha}(i)\tag{A14}\\
&  =\sum_{j\in T}b_{\alpha}(j)=-\sum_{j\in\overline{T}}b_{\alpha}(j);\nonumber
\end{align}
which tells that the taxicab principal dimensions are \textit{balanced}; and%

\begin{align}
\delta_{\alpha}/4  &  =\sum_{(i,j)\in S\times T}X_{\alpha}(i,j)=\sum
_{(i,j)\in\overline{S}\times\overline{T}}X_{\alpha}(i,j)\tag{A15}\\
&  =-\sum_{(i,j)\in\overline{S}\times T}X_{\alpha}(i,j)=-\sum_{(i,j)\in
S\times\overline{T}}X_{\alpha}(i,j),\nonumber
\end{align}
which tells that the $\alpha$-th principal dimension divides the residual data
matrix $\mathbf{X}_{\alpha}$ into 4 \textit{balanced} quadrants, see
Choulakian and Abou-Samra (2020).

\textit{Step 3}: Calculate taxicab principal factor scores ($f_{\alpha
}(i),g_{\alpha}(j))\ $of $\mathbf{X}$ by dividing each term in (13) by the
weights $m_{i}^{r}m_{j}^{c}$%
\begin{equation}
y_{ij}=\sum_{\alpha=1}^{k}f_{\alpha}(i)g_{\alpha}(j)/\delta_{\alpha},
\tag{A16}%
\end{equation}
where evidently $f_{\alpha}(i)=$ $a_{\alpha}(i)/m_{i}^{r}$ and $g_{\alpha
}(j)=b_{\alpha}(j)/m_{j}^{c}.$ Equation (A16) is named \textquotedblright data
reconstruction formula\textquotedblright.

The principal maps are obtained by plotting ($f_{1}(i),f_{2}(i))$ and
($g_{1}(j),g_{2}(j)),$\bigskip
\begin{verbatim}
Appendix 3: QSR index
\end{verbatim}

Let
\[
QSR_{\alpha}=\frac{\delta_{\alpha}}{\sum_{(i,j)}|X_{\alpha}(i,j)|},
\]
which we will interpret as a new intrinsic measure of quality of the signs of
the residuals in the residual matrix $\mathbf{X}_{\alpha}$ for $\alpha
=1,...,k$.

Let $S\cup\overline{S}=I$ be the optimal principal axis partition of $I$, and
similarly $T\cup\overline{T}=J$ be the optimal principal axis partition of
$J,$ such that $S=\left\{  i:a_{\alpha}(i)>0\right\}  =$ $\left\{
i:v_{\alpha}(i)>0\right\}  $ and $T=\left\{  j:b_{\alpha}(j)>0\right\}
=\left\{  j:u_{\alpha}(j)>0\right\}  $ by (A4). Thus the data set is divided
into 4 quadrants. Based on the equations (A15), we define a new index
quantifying the quality of the signs of the residuals in each quadrant of the
$\alpha$th residual matrix $\mathbf{X}_{\alpha}$ for $\alpha=1,...,k$ to be
\bigskip%
\begin{align*}
QSR_{\alpha}(E,F)  &  =\frac{\sum_{(i,j)\in E\times F}v_{\alpha}(i)X_{\alpha
}(i,j)u_{\alpha}(j)}{\sum_{(i,j)\in E\times F}|X_{\alpha}(i,j)|}\\
&  =\frac{\delta_{\alpha}/4}{\sum_{(i,j)\in E\times F}|X_{\alpha}(i,j)|}\text{
for }(E,F)=(\overline{S},\overline{T})\text{ or }(S,T)\\
&  =\frac{-\delta_{\alpha}/4}{\sum_{(i,j)\in E\times F}|X_{\alpha}%
(i,j)|}\text{ for }(E,F)=(S,\overline{T})\text{ or }(\overline{S},T)
\end{align*}
Sometimes we express it also in \%.\bigskip

We have the following easily proved\bigskip

\textbf{Lemma }: a) $-1\leq QSR_{\alpha}(E,F)\leq1;$ furthermore,
$QSR_{\alpha}=1$ if and only if $QSR_{\alpha}(S,T)=QSR_{\alpha}(\overline
{S},\overline{T})=-QSR_{\alpha}(S,\overline{T})=-QSR_{\alpha}(\overline
{S},T)=1,$ for $\alpha=1,...,k-1.$

b) For $\alpha=k,$ $QSR_{\alpha}=1.\bigskip$

The interpretation of $QSR_{\alpha}(E,F)=\pm1$ is that in the quadrant
$E\times F$ the residuals have one sign; and this is a signal for very
influential cells or columns or rows; for an example see Choulakian
(2021).\bigskip

\textbf{References}\bigskip

Aitchison J (1983) Principal component analysis of compositional data.
\textit{Biometrika} 70(1):57--65

Aitchison J (1986) \textit{The Statistical Analysis of Compositional Data}.
London: Chapman and Hall notes---monograph series, Institute of Mathematical
Statistics, Hayward, 73--81

Aitchison J (1994) Principles of compositional data analysis. A chapter in
\textit{Multivariate analysis and its applications}, volume 24 of lecture

Aitchison J (1997) \textit{The one-hour course in compositional data analysis
or compositional data analysis is simple}. In: Pawlowsky-Glahn V(ed)
Proceedings of IAMG'97---the III annual conference of the international
association for mathematical geology, volume I, II and addendum, Barcelona
(E). CIMNE, Barcelona, pp 3--35, ISBN 978-84-87867-76-7

Allard J, Champigny S, Choulakian V, Mahdi S (2020) TCA and TLRA: A comparison
on contingency tables and compositional data.Available
at\ \textit{https://arxiv.org/pdf/2009.05482.pdf}

Benz\'{e}cri JP (1973a).\ \textit{L'Analyse des Donn\'{e}es: Vol. 2: L'Analyse
des Correspondances}. Paris: Dunod

Choulakian V (2006) Taxicab correspondence analysis. \textit{Psychometrika,}
71, 333-345

Choulakian V (2016) Matrix factorizations based on induced norms.
\textit{Statistics, Optimization and Information Computing}, 4, 1-14

Choulakian V (2021) Quantification of intrinsic quality of a principal
dimension in correspondence analysis and taxicab correspondence analysis.
Available on \textit{arXiv:2108.10685}

Choulakian V (2022) Some notes on Goodman's marginal-free correspondence
analysis. \textit{https://arxiv.org/pdf/2202.01620.pdf}

Choulakian V, Abou-Samra G (2020) Mean absolute deviations about the mean, the
cut norm and taxicab correspondence analysis. \textit{Open Journal of
Statistics}, 10(1), 97-112

Choulakian V, Allard J, Smail M (2023) Taxicab Correspondence Analysis and
Taxicab Logratio Analysis: A Comparison on Contingency Tables and
Compositional Data. To appear in the \textit{Austrian Journal of Statistics}

D'Ambra L, Amenta P, D'Ambra A, De Tibeiro J (2020) A study of the family
service expenditures and the socio-demographic characteristics via fixed
marginals correspondence analysis. \textit{Socio-Economic Planning Sciences},
73(1). DOI:10.1016/j.seps.2020.100833

Goodman LA (1996). A single general method for the analysis of
cross-classified data: Reconciliation and synthesis of some methods of
Pearson, Yule, and Fisher, and also some methods of correspondence analysis
and association analysis.\textit{\ Journal of the American Statistical
Association}, 91, 408-428.

Greenacre M, Lewi P (2009) Distributional equivalence and subcompositional
coherence in the analysis of compositional data, contingency tables and
ratio-scale measurements. \textit{Journal of Classification}, 26, 29-54.

Lauro NC, D'Ambra L (1984) L'analyse non symetrique des correspondances. In
\textit{Data Analysis and Informatics III} (eds. E Diday et al.), 433-446,
Amsterdam: Elsevier.

Pawlowsky-Glahn V, Egozcue JJ (2011) Exploring compositional data with the
CoDa-Dendrogram. \textit{Austrian Journal of Statistics, }40(1\& 2), 103-113

Yule GU (1912) On the methods of measuring association between two attributes.
\textit{JRSS}, 75, 579-642

\end{document}